\documentclass{icrc2009}

\usepackage{graphicx}
\usepackage[caption=false]{caption}
\usepackage[font=footnotesize]{subfig}
\usepackage{fixltx2e}
\usepackage{url}

\newcommand{\shorttitle}[1]
{\markboth{Proceedings of the 31\MakeLowercase{$^{st}$} ICRC, {\L}\'{o}d\'{z} 2009}{#1} }
\newcommand{\etal}{\MakeLowercase{\textit{et al. }}}

\begin{document}
\title{VERITAS observations of M87 from 2007 to present}

\author{\IEEEauthorblockN{Chiumun Michelle Hui\IEEEauthorrefmark{1}
			  for the VERITAS collaboration\IEEEauthorrefmark{2}}
                            \\
\IEEEauthorblockA{\IEEEauthorrefmark{1}University of Utah, Department of
  Physics and Astronomy, Salt Lake City, Utah 84112, USA (cmhui@physics.utah.edu)}
\IEEEauthorblockA{\IEEEauthorrefmark{2} see R. A. Ong et al (these
  proceedings) or http://veritas.sao.arizona.edu/conferences/authors?icrc2009 }} 

\shorttitle{C. M. Hui \etal VERITAS observations of M87}
\maketitle

\begin{abstract}
M87 is a nearby radio galaxy and because of its misaligned jet, it is possible
to correlate detailed spatially-resolved emission regions in the radio,
optical to X-ray waveband with unresolved but contemporaneous flux
measurements in the TeV regime. Hence, M87 provides a unique opportunity to
reveal the emission mechanisms responsible for high energy gamma-ray emission
from active galactic nuclei. Observations with VERITAS since 2007 have
resulted in 90 hours of data while 2008 observations were part of a concerted
effort involving the three major atmospheric Cherenkov observatories:
H.E.S.S., MAGIC and VERITAS. As a result of the TeV campaign, a high flux
state of M87 was detected in February 2008 showing multiple flares with rapid
variability. We will present the comprehensive results from VERITAS
observations since 2007 and also show preliminary results from the 2009
campaign.
\end{abstract}


\section{Introduction}
M87 is a giant elliptical galaxy located 16\,Mpc away (redshift
$\rm{z}=0.00436$) near the center of the Virgo cluster.  It has been observed at
all wavelengths ranging from radio to TeV gamma rays. Its core is an active
galactic nucleus (AGN) powered by a supermassive black hole of $\sim 3.2
\times 10^9\,\rm{M}_\odot$ \cite{macchetto97}.  The jet of M87 does not point
along our line of sight; however apparent superluminal motion has been
observed in both radio \cite{cheung07} and optical \cite{biretta99} for
different features along the jet, constraining the jet orientation to
$<30^\circ$ at the location of the knot HST-1.  M87 is described as a
misaligned BL Lac \cite{tsvetanov}.  The proximity of M87 and its misaligned
jet have enabled the study of its jet morphologies, which are similar in
radio, optical, and X-rays \cite{perlman05}.  Flaring activities have been
observed at these energies simultaneously and in different jet features
\cite{cheung07}, which revealed many characteristics of relativistic jets in
AGN.

TeV emission from M87 was discovered by the HEGRA collaboration from their
1998-1999 observations \cite{hegra03} and was confirmed by the
H.E.S.S. collaboration \cite{hess06}, which additionally reported year-scale
and days-scale flux variability during a high state of gamma-ray activity in
2005.  The observed variability timescales disfavor large scale gamma ray
production models such as the dark matter annihilation model \cite{baltz99}
and the interacting cosmic ray proton scenario \cite{pfrommer03}, and favor
the immediate vicinity of the M87 black hole as the TeV production site.
However, the angular resolution of imaging atmospheric Cherenkov telescopes
(IACTs) is insufficient to resolve any structure in M87.  During the same
period of the flare observed by H.E.S.S. in 2005, the Chandra X-ray
observatory detected the knot HST-1 ($\sim 0.8$'' away from the nucleus) at an
intensity more than 50 times that observed in 2000.  HST-1 was then suggested
as a more likely source of TeV emission than the core \cite{harris08}.

The knot HST-1 has been demonstrated as a possible location for jet
reconfinement where photons can be upscattered to TeV energies via the
inverse-Compton process \cite{stawarz06}.  Several models with emission
originating at the inner jet region have also been proposed, where TeV emission
can be produced via inverse-Compton scattering \cite{geor05} or via synchrotron
self-Compton processes involving more complex jet structures \cite{lenain08} 
\cite{tavecchio08}.  Leptonic models involving the electromagnetic field of
the black hole \cite{neronov07} \cite{rieger08} with TeV emission coming from
the vicinity of the black hole and not the inner jet have also been suggested.

VERITAS confirmed TeV emission above 250\,GeV from M87 in the 2007 dataset, but
at a lower flux than what was reported by H.E.S.S. in 2005, and no variability
was detected for 2007 \cite{veritas08}.  Day-scale variability was reported in
2008 by the MAGIC collaboration \cite{magic08} with a 13-day flare
during the 2008 TeV joint monitoring campaign by H.E.S.S., MAGIC,
and VERITAS \cite{joint}.  VERITAS observation was intensified following a
trigger alert issued by the MAGIC collaboration and another flare was detected
\cite{veritas09}.  Chandra X-ray data taken in the same month (6-week sampling
frequency) showed historical maximum activity coming from the core while the
nearby knot HST-1 remained quiescent \cite{harris09}.  The contemporaneous
gamma-ray and X-ray flares suggest the core as the more probable TeV emission
region, in contrast to the 2005 flaring activity in TeV observed by
H.E.S.S. and in the knot HST-1 in X-rays by Chandra.  Publication for the results from this joint campaign is forthcoming \cite{joint2}.

In 2009 the TeV monitoring campaign is continued with MAGIC and VERITAS.  In
this paper, we will present preliminary results of the VERITAS 2009 dataset,
along with the results from observations beginning in 2007.

\section{VERITAS observations}
VERITAS, the Very Energetic Radiation Imaging Telescope Array System, is an
array of four 12\,m diameter imaging atmospheric Cherenkov telescopes located
at the Fred Lawrence Whipple Observatory at Mount Hopkins in southern Arizona.
Each telescope is equipped with a camera comprising 499 photomultiplier tubes
arranged in a hexagonal lattice covering a field of view of $3.5^\circ$.  The
array is sensitive from 100\,GeV to more than 30\,TeV.  It has an effective
area of $\sim 10^5\,\rm{m}^2$ and an angular resolution of $\sim 0.1^\circ$
($68\%$ containment).  For more details of VERITAS, see \cite{acciari08}.  

M87 was observed with VERITAS for over 115 hours between February 2007 and April
2009 at a range of zenith angles from $19^\circ$ to $41^\circ$.  Observations
in spring 2007 were carried out during the construction phase and only
3-telescope data (94\% of spring 2007 data) are used in the spectral analysis.
Later observations (fall 2007 onward) were achieved with 4 telescopes.  All
observations were performed in wobble mode where M87 is tracked with a
$0.5^\circ$ offset to the camera center.  After eliminating bad weather 
observations and unstable trigger rate data, over 90 hours of quality live data
were then processed with several independent analysis packages \cite{daniel07}
with slightly different algorithms.  All analysis packages yield consistent
results. 

Shower images are first corrected in gain and timing using parameters obtained
from the nightly laser calibration data. Then the images are passed through a
two-threshold cleaning.  Each shower image is then parametrized
\cite{hillas85}, and the shower direction is reconstructed using the
stereoscopic technique.  Events are selected as gamma-ray like if at least two
images passed cuts optimized for a $10\%$ Crab Nebula flux source.  The source
region is defined by a $0.15^\circ$ radius disk centered on the source
coordinates, and all the gamma-ray like events within this region are summed
to the ON count; the background is estimated from seven identically sized
regions reflected from the source region around the camera center, and is
summed to the OFF count \cite{berge07}.  The ON and OFF counts are then used
in the Li \& Ma formula 17 \cite{LiMa83} to calculate the significance of the
excess.

\section{Results}
In 2007, M87 was detected at a statistical significance of $5.9\,\sigma$ after
44 hours of observations between February and April with a 3-telescope array.
An average flux of $(3.47 \pm 1.12) \times 10^{-12} \rm{cm}^{-2} \rm{s}^{-1}$
for energies above 250\,GeV was measured from this dataset, corresponding to
$\sim2\%$ of the Crab Nebula flux.  No significant short-term flux variability
was detected \cite{veritas08}. 

In 2008, M87 was detected at $7.2\,\sigma$ after 41 hours of observations
between December 2007 and May 2008.  An average flux of $(2.74 \pm 0.93)
\times 10^{-12}\rm{cm}^{-2}\rm{s}^{-1}$ above 250\,GeV was recorded,
corresponding to $\sim2\%$ of the Crab Nebula flux.  During 4 days in February
2008 (MJD 54505 - 54509) flaring activity was observed in 6 hours of data,
resulting in a $7.4\,\sigma$ detection and an average flux that corresponded
to 5.3\% of the Crab Nebula flux.  15 hours of observations were performed
before the flare period and resulted in a marginal detection ($<5\,\sigma$) of
M87 at 2.0\% of the Crab Nebula flux; after the flare period, 19 hours of
observations yielded no detection of M87 and an upper limit of $<1.4\%$ of the
Crab Nebula flux.  The spectral index ($\Gamma$) of the differential spectrum
power-law fit with the form $d\Phi/dE = \Phi_0 (E/TeV)^{-\Gamma}$ showed no
significant variation between pre-flare and flare period (see Table
\ref{tabspec}).  

\begin{figure*}[t]
  \centering
  \includegraphics[width=\textwidth]{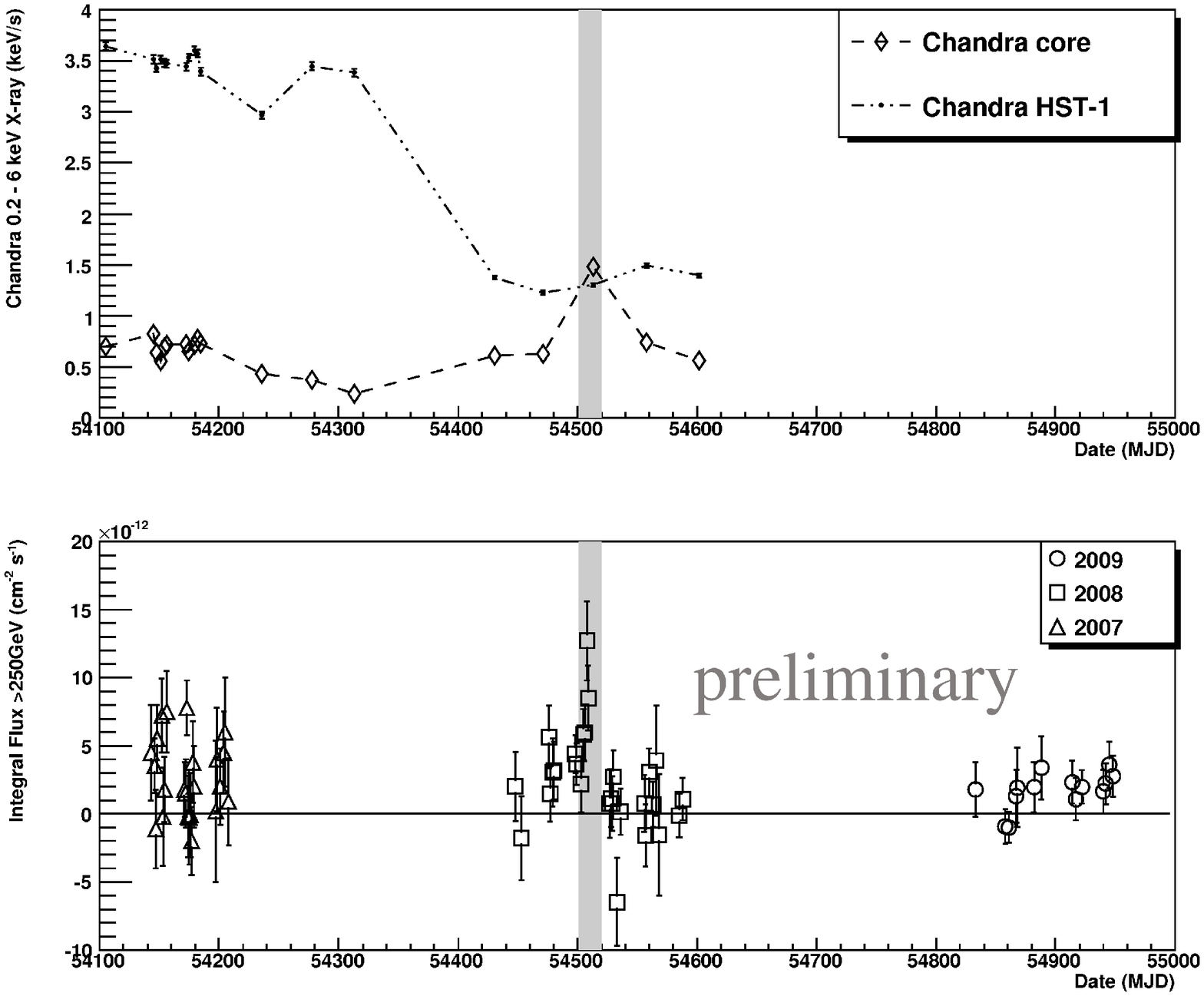}
  \caption{\emph{Upper panel:} Chandra X-ray lightcurves measured from the
    core and the knot HST-1.  \emph{Lower panel:} Nightly fluxes for energies
    above 250\,GeV from 2007 to present.  Grey area highlights the 2008 flare
    period observed by VERITAS and the corresponding X-ray fluxes observed by Chandra.  For details on the 2008 flare period, see \cite{joint2}}
  \label{lc}
\end{figure*}

In 2009, 18 hours of observations between January and April yielded a marginal
detection of M87 and an upper limit of $<1.9\%$ Crab Nebula flux.  No flaring
activity was observed (see figure \ref{lc}). 

\begin{table}[!h]
  \caption{Differential spectrum power-law fit of the form $d\Phi/dE = \Phi_0
    (E/TeV)^{-\Gamma}$} 
  \label{tabspec}
  \centering
  \begin{tabular}{|c|c|c|}
  \hline
   data  & $\Phi_0$  & $\Gamma$ \\
   & ($10^{-13} cm^{-2} s^{-1} TeV^{-1}$) & \\
   \hline 
    2007 & $7.4 \pm 1.3$ & $2.31 \pm 0.17$ \\
    2008-all & $5.2 \pm 0.9$ & $2.49 \pm 0.19$ \\
    2008-flare & $15.9 \pm 2.9$ & $2.40 \pm 0.21$ \\
    2008-preflare & $5.6 \pm 1.5$ & $2.49 \pm 0.26$ \\
  \hline
  \end{tabular}
\end{table}

\begin{figure}[h]
  \centering
  \includegraphics[width=0.5\textwidth]{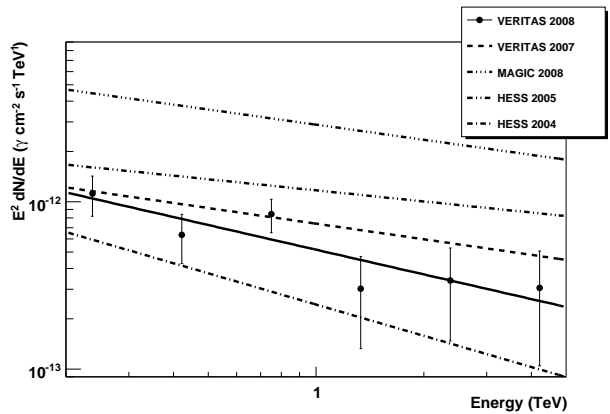}
  \caption{VERITAS M87 energy spectrum of the entire dataset, in comparison to
    TeV energy spectra reported in the past.  The spectral index of all datasets
    ranged from 2.22 to 2.62, and are compatible within statistical errors.} 
  \label{spec}
\end{figure}

\section{Discussion}
VERITAS observations of M87 spanning three observing seasons (from 2007 to
present) have shown M87 in steady emission state and flaring state.  The
spectra obtained from these observations show no significant changes in the
spectral index.  During the 2008 joint monitoring campaign of M87, VERITAS
observed a gamma-ray flare in February 2008 which spanned 4 days, constraining
the emission region size to $R \le R_{var} = \delta \, c\, \Delta t /(1+z) =
\delta\,10^{16} \rm{cm} \sim 11.1\delta R_s$ where $\delta$ is the relativistic
Doppler factor and $R_s$ the Schwarzschild radius of the M87 black hole.
Rapid variability reported previously \cite{hess06} \cite{magic08} constrained
the size of the TeV emission region to $< 2.6\delta R_s$.

Even though the gamma-ray observing technique cannot resolve individual features
of M87, the TeV emission size constraint has narrowed down the most probable TeV
emission location to the unresolved core region and the knot HST-1.  The 2008
gamma-ray flare coincided with the Chandra observation of historically high flux
coming from the core while the nearby knot HST-1 appeared to be inactive
(figure \ref{lc}) \cite{joint2}.  The contemporaneous gamma-ray and X-ray flares suggest the core is more likely the TeV emission region, in contrast to the 2005 flaring activity in TeV observed by H.E.S.S. and in the knot HST-1 in X-rays by Chandra \cite{harris08}.

The 2008 multi-wavelength observations of M87 included concurrent radio,
X-ray, and TeV gamma-ray coverage of M87 flaring activity from the core region
in early 2008 \cite{wagner} \cite{joint2}.  From the 2008 multiwavelength data, the TeV
emission region is likely the unresolved core.  However, the knot HST-1 is
still a possible candidate for the 2005 flare.  Current models do not favor
one over the other, and both the core and HST-1 remain as candidates for TeV
emission.  As of the end of April 2009, the 2009 monitoring work has shown no
flaring activity from M87.  Further multi-wavelength monitoring can
potentially provide additional constraints on the environment of M87 and more
insights into the emission mechanism of AGN.

\section*{Acknowledgments}
This research is supported by grants from the US Department of
Energy, the US National Science Foundation, and the Smithsonian
Institution, by NSERC in Canada, by Science Foundation Ireland, and
by STFC in the UK. We acknowledge the excellent work of the technical
support staff at the FLWO and the collaborating institutions in the
construction and operation of the instrument.

\end{document}